# Mössbauer measurements of GaFeO$_3$ single crystal multiferroic

K. Szymański[1], W. Olszewski[1], P. Butkiewicz[1], M. Biernacka[1], D. Satuła[1], K. Rećko[1], M. Szafrański[2]

[1]Faculty of Physics, University of Bialystok, K. Ciolkowskiego 1L,
15–245 Bialystok, Poland
[2]Faculty of Physics, Adam Mickiewicz University, Umultowska 85, 61-614 Poznan, Poland

email k.szymanski@uwb.edu.pl

abstract

Mössbauer measurements on single crystal absorbers at room and at low temperatures were performed. The results are fully consistent with previously published reports by other groups. Spectra of single crystals were simultaneously analyzed including magnetic dipole and electric quadrupole interactions. The analysis shows that there is a small component of magnetic moments perpendicular to the magnetic easy axis. Mössbauer data seem not agree with commonly accepted ferrimagnetic structure of GaFeO$_3$.

## 1. Introduction

Recently there has been a rise of interest in multiferroics, single phase compounds in which magnetism and ferroelectricity coexist. One example is GaFeO$_3$ [1] in which linear magnetoelectric effect [2], large magnetic anisotropy [3], and magnetization with concentration controlled transition temperature [4] were observed. The compound crystallizes in the orthorhombic structure of space group $Pc2_1n$ [5], [6], [7], [8], [9]. The unit cell contains two nonequivalent gallium (Ga$_1$, Ga$_2$), two iron (Fe$_1$, Fe$_2$) and six oxygen sites. The cation in Ga$_1$ has tetrahedral coordination while in remaining Ga$_2$, Fe$_1$ and Fe$_2$ cations are octahedrally coordinated. Pronounced cation site disorder between Ga$_2$ and Fe$_1$, Fe$_2$ exists [5], [6], [8]. The different sample preparation methods, influencing mainly features of the Ga/Fe site disorder results in different magnetic transition temperatures – floating-zone method: 200 [6], flux growth: 260 [1], 285 [4], 260 [10], 255 [11], 292 [12]; solid state reaction: 210 [13], 165 [14] (all in K). The Fe atoms on Fe$_2$ site and those on Ga$_2$ site form up magnetic moments, whereas those on Fe$_1$ site form down-moments resulting in ferrimagnetic structure [6], [12], [13], [15], [16]. The net magnetic moment about 0.7 $\mu_B$/Fe at ground state is result of site disorder. The ionic state of iron is Fe$^{3+}$ with a half-filled $3d^5$ configuration, in which the orbital momentum vanishes. Despite of it magnetocrystalline anisotropy as large as 1.1 was measured [17] and around 2 [18], [19] calculated (in meV per unit cell). It was found by X-ray magnetic dichroism that iron atoms have orbital momentum $L=0.23\pm0.04$ caused by anisotropic O$2p$-Fe$3d$ hybridization [10]. Electrical polarization, is difficult for observation in bulk crystals because of large leakage currents. Theoretical calculations predict value of spontaneous electrical polarization [20], [21], close to the value estimated by a point charge approximation for perfectly ordered GaFeO$_3$. However, the mechanism of polarization switching is still under debate. Also, the observed electrical polarization switching in Cr doped GaFeO$_3$ films [22] seems to be inconsistent with the switching mechanism proposed in [23], [21]. Magnetoelectric effect, resulting in coupling of magnetization and electrical polarization is studied intensively. The antisymmetric components of the magnetoelectric effect tensor can couple also to the ferrotoroidic order parameter [24] and this order was detected in GaFeO$_3$ by direction dependent scattering of X-rays [25]. In this context details of local magnetic and electronic structure are highly required. Local structure was investigated in Mössbauer experiments on polycrystalline samples [15], [26], [14], [27] also in high magnetic field [11], [13], [28]. In experiments on single crystals hyperfine structure was interpreted by calculations of the electric field gradient in point charge model [29], [30]. There is puzzling inconsistency between almost vanishing line intensities no 2 and 5 for $Ga_{0.85}Fe_{1.15}O_3$ [26] and non-negligible value in $GaFeO_3$ [30] indicating possible magnetic component perpendicular to the magnetization easy axis. In our paper preliminary results of the Mössbauer measurements performed on the oriented single-crystal absorbers are presented. The orientations of hyperfine interactions are presented.

## 2. Sample preparation

$GaFeO_3$ single crystals were grown by the optical floating zone crystal growth technique following previous approaches [6], [8], [31]. The starting materials were $Fe_2O_3$ (99.999%, Acros Organics) and $Ga_2O_3$ (99.99%, Sigma Aldrich) powders. Stoichiometric amounts of the oxides were thoroughly grounded together, pre-calcined in a horizontal furnace tube at 1100°C for 10 h and calcined at 1250°C over next 10 h in the air, with intermediate regrinding. Then the powder was loaded into the rubber tubes and compacted using a hydraulic press at a pressure of 70 MPa. The obtained rods were sintered for 72 h in air at 1385°C in a vertical furnace tube and then cooled to the room temperature at a rate of 1°C/min. Single crystals growth was performed with a four-mirror optical floating zone furnace (FZ-T-4000-H, Crystal Systems Corp. Japan, lamps power 4x300 W) in pure oxygen.

One of the crystal growth procedure was designed so, to have electrical polarization axis parallel to the crystal growth direction. To achieve the goal a seed rod with appropriate orientation was prepared. At the beginning a polycrystalline rod was used as a seed and a necking technique was applied during growth. That way a single crystal was obtained by spontaneous nucleation. The crystal was next cut to form a few millimetres high pyramid with appropriate orientation. The base of the pyramid was attached with PVA binder to the top of the flat polycrystalline rod. The rod with the pyramid on the top was sintered at 1380°C in the air by 48 h and slowly cooled to RT resulting in strong enough joint. Another successful reinforced direction of the crystal growth was along direction of the magnetic easy axis.

## 3. Sample characterization

The quality of the single crystals and their composition were checked by X-ray diffraction. A small piece, cut off from the large crystal, was used for this purpose. The studies were performed on an Oxford Diffraction Gemini A Ultra diffractometer operating with graphite-monochromated Mo$K_\alpha$ radiation. CrysAlis$^{Pro}$ software [32] was used for data collection and processing. The crystal structure was solved with direct methods using SHELXS-97 and refined by full-matrix least-squares method on all intensity data with SHELXL-97 [33]. Independent structural characterization of the powder obtained by crushing the single crystals was carried out using an Empyrean PANalytical powder diffractometer in Bragg-Brentano geometry with Mo$K_\alpha$ radiation. The diffraction patterns were analyzed by Rietveld – type profile refinement method using FullProf program [34]. The site occupancies were implemented from the Mössbauer spectroscopy data (see next section). The obtained structural parameters are consistent with those reported in [6], [8], [9].

The temperature of magnetic phase transition obtained from an inflection point of FC curve for magnetic field of 0.05 T applied in *c* direction is 201(5) K. The *M(H)* measurements yield a spontaneous magnetization value at 2 K equal 0.74 $\mu_B$/Fe, which is slightly more than value of 0.67 $\mu_B$/Fe at 5 K reported in [6].

### 4. Mössbauer measurements

Mössbauer measurements were performed using the spectrometer operating in a constant acceleration mode and a $^{57}$Co source in a Rh matrix. The velocity scale was calibrated using α-Fe standard foil also at room temperature. Closed cycle refrigerator equipped in antivibrational shroud was used for low temperature measurements. The single crystal rod was cut to slices, glued to plexi plate and mechanically thinned to appropriate for measurements thickness between 20 and 70 μm. Two absorbers were prepared, one with the easy magnetization axis, the other one with the electrical polarization direction perpendicular to the sample plane.

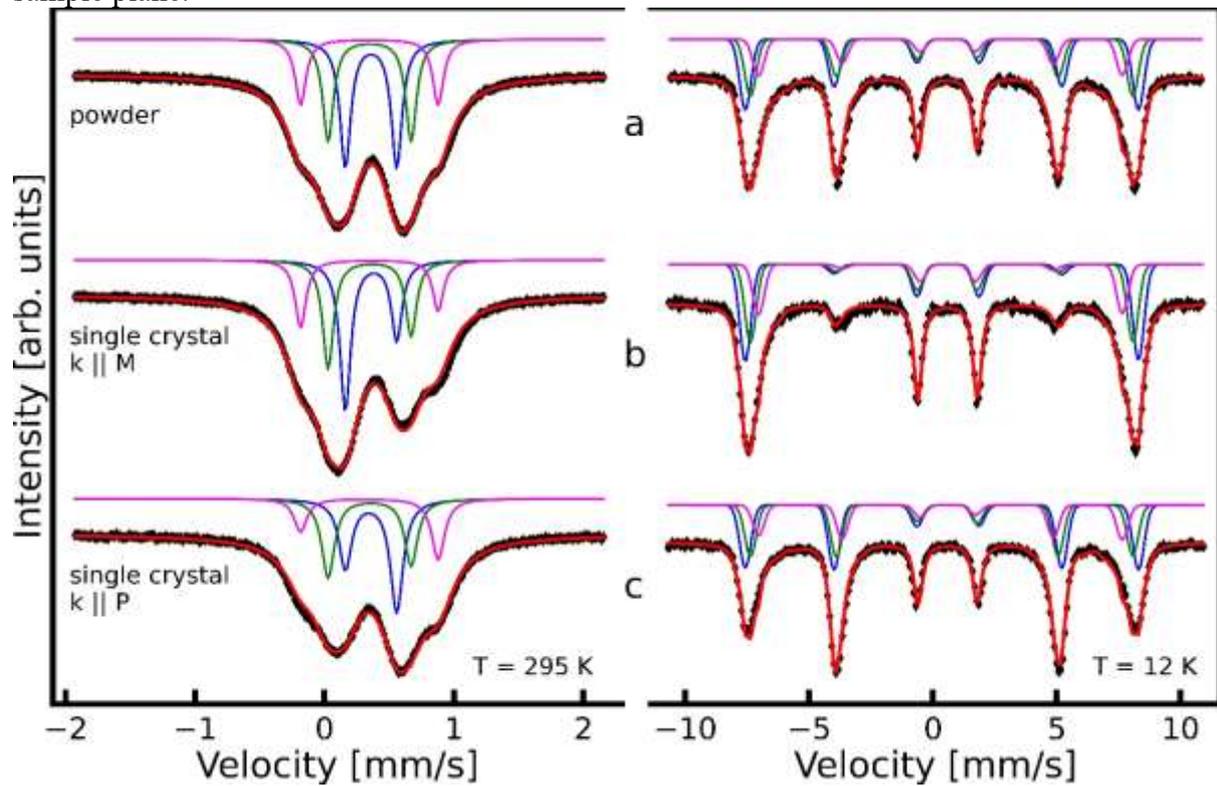

Fig. 1 Mössbauer spectra of a) powder and oriented single crystals with **k** vector parallel to the b) magnetization easy axis and c) electrical polarization. Lines with green, blue and magenta colours correspond to the iron at sites $Fe_1$, $Fe_2$, $Ga_2$, respectively. Details of spectra analysis are given in next section.

The measured spectra are fully consistent with earlier published data on single crystals [29], [30]. There are clear nonvanishing line intensities at velocities -4 and 5 mm/s in Fig. 1b. In order to check whether nonvanishing lines may be result of nonperfect crystal orientation, we have performed additional measurements (Fig. 2) with **k** vector of photons inclined from the normal to the sample plane directions, as shown in Fig. 3.

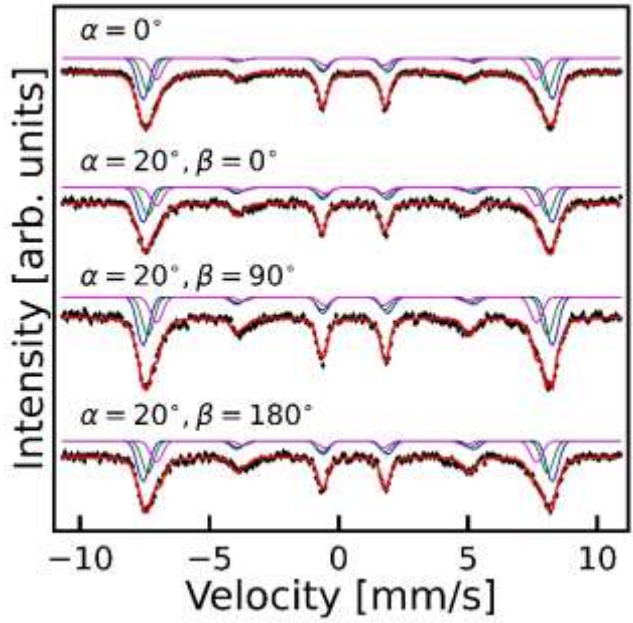

Fig. 2. Mössbauer spectra measured with ***k***-vector directions inclined from the normal to the sample plane The definitions of angles $\alpha$, $\beta$ are shown in Fig. 3.

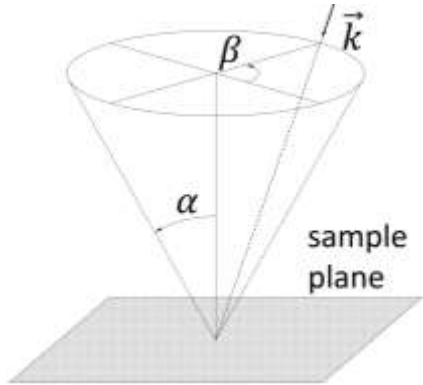

Fig. 3 Schematic arrangement of sample plane and k-vector of photon.

5. **Data analysis and discussion**

The Mössbauer spectra were analysed by commercially available NORMOS package [35] within the transmission integral approach. Three components, corresponding to iron atoms at $Fe_1$, $Fe_2$ and $Ga_2$ crystallographic positions, were fitted to the spectrum. Components measured at room temperature consist of two lines (quadrupole interactions only) while those measured at low temperatures with eight lines (mixed dipolar magnetic and quadrupole electric); positions and intensities of the absorption lines were calculated in the formalism of the intensity tensor [36], [37]. The way of calculations the line intensities and positions for spectra of $GaFeO_3$ single crystals were presented in [29], [30]. We use equivalent approach based on algebraical method.

Atom located at position $(x, y, z)$ of the $GaFeO_3$ unit cell interact with the electric field gradient (EFG) tensor and its principal axis system consist of three orthogonal unit vectors $\boldsymbol{e}_1, \boldsymbol{e}_2, \boldsymbol{e}_3$. Cartesian components of the $\boldsymbol{e}_1$ in orthogonal unit frame of main crystallographic directions are: $\boldsymbol{e}_1 = (e_{1x}, e_{1y}, e_{1z})$, and so for the $\boldsymbol{e}_2$ and $\boldsymbol{e}_3$. In the $Pc2_1n$ space group operations transforming any point $(x, y, z)$ into $(1/2 - x, y, 1/2 + z)$, $(-x, 1/2 + y, -z)$ or $(1/2 + x, 1/2 + y, 1/2 - z)$ are symmetries of the structure and they are the glide $yz$-plane, two fold $y$-screw axis and glide $xy$-plane, respectively. A components $(e_{1x}, e_{1y}, e_{1z})$ of the vector $\boldsymbol{e}_1$, transform thus into $(-e_{1x}, e_{1y}, e_{1z})$, $(-e_{1x}, e_{1y}, -e_{1z})$ and $(e_{1x}, e_{1y}, -e_{1z})$ under the glide *yz*-plane, the two fold *y*-screw axis and the glide *xy*-plane, respectively. Vectors $e_2$ and $e_3$ are transformed similarly and the elements of the point group symmetries are diagonal matrices $P_i$, $i = 1...4$ with diagonal elements $(1,1,1)$, $(-1,1,1)$, $(-1,1,-1)$, $(1,1,-1)$, respectively. Writing three vectors of a reference frame

$$R_1 \stackrel{\text{def}}{=} [e_1, e_2, e_3] = \begin{bmatrix} e_{1x} & e_{1y} & e_{1z} \\ e_{2x} & e_{2y} & e_{2z} \\ e_{3x} & e_{3y} & e_{3z} \end{bmatrix}, \quad (1)$$

one express components of transformed frames $R_i$ as product of matrices

$$R_i = P_i \cdot [e_1, e_2, e_3] \cdot P_i \quad (2)$$

There are four different local orientations of the principal axes of the EFG for each atomic position. To illustrate it, we show as an example four $Fe_1$ atoms, having these four different orientations of local environment, labelled by $Fe_1^{(i)}, i = 1...4$ in Fig. 4. The different orientations are illustrated further in Fig. 4, where by red, green and blue bars three shortest $Fe_1 - O$ bonds are represented. Orientation of the bonds are related to local orientation of the principal axes of the EFG. The handedness of local environments $Fe_1^{(1)}$ and $Fe_1^{(3)}$ in Fig. 4 is opposite to that of $Fe_1^{(2)}$ and $Fe_1^{(4)}$, which is result of $P_i$ symmetries.

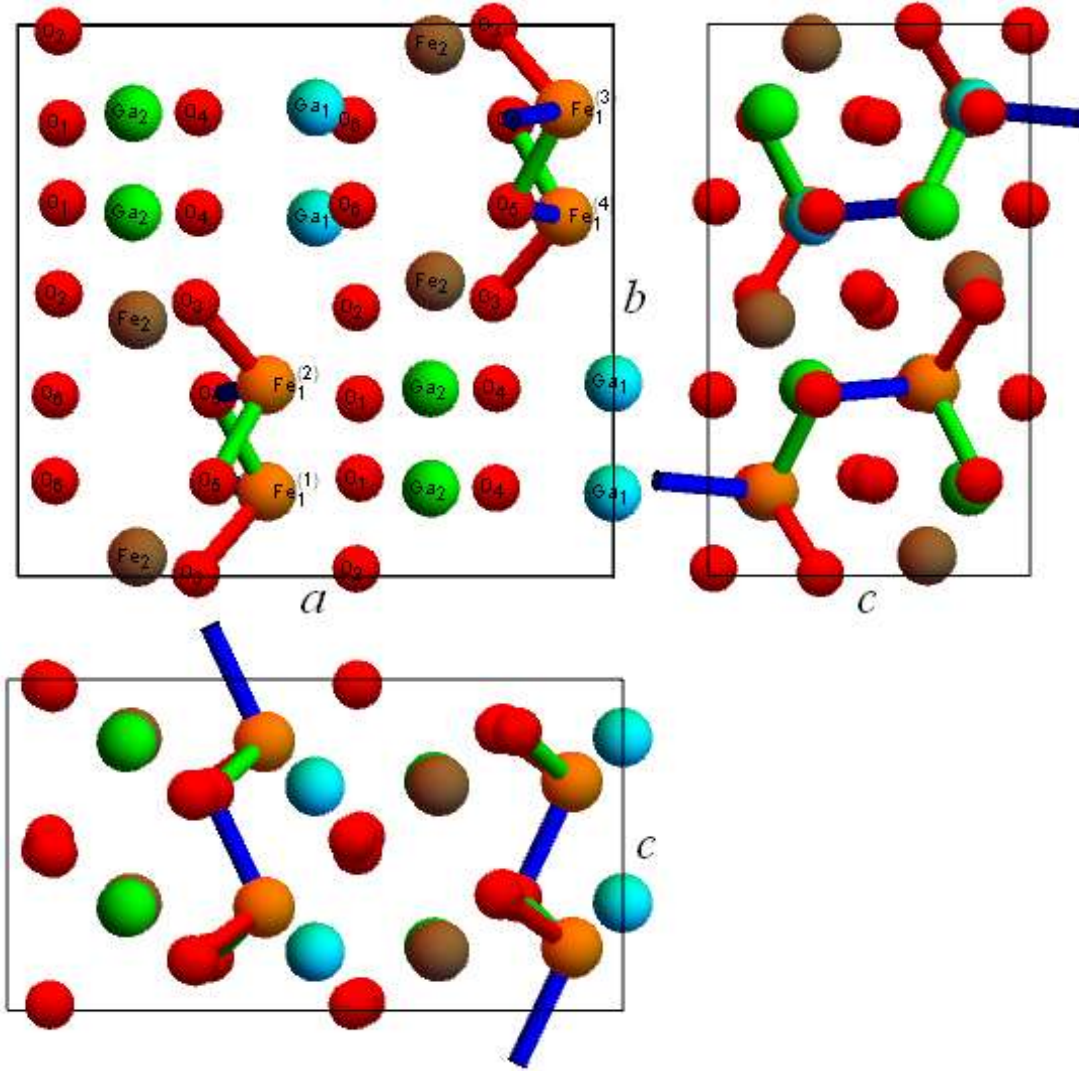

Fig 4. Projection of the crystal structure of GaFeO$_3$ along the crystal axes (space group *Pc2$_1$n*). By red, green and blue bars the shortest Fe$_1$-O bonds are shown which may be used for orientation of a local reference frames.

A convenient, parametrization of the $e_1, e_2, e_3$ vector components (1) of local frames is

$$\begin{aligned}
\boldsymbol{e}_1 &= (\cos\theta\cos\varphi\cos\psi - \sin\varphi\sin\psi, -\cos\theta\cos\varphi\sin\psi - \sin\varphi\cos\psi, \sin\theta\cos\varphi), \\
\boldsymbol{e}_2 &= (\cos\theta\sin\varphi\cos\psi + \cos\varphi\sin\psi, -\cos\theta\sin\varphi\sin\psi + \cos\varphi\cos\psi, \sin\theta\sin\varphi), \\
\boldsymbol{e}_3 &= (-\sin\theta\cos\psi, \sin\theta\cos\psi, \cos\theta),
\end{aligned} \qquad (3)$$

because in the $\boldsymbol{e}_1, \boldsymbol{e}_2, \boldsymbol{e}_3$ local frame the unit vector $\boldsymbol{z} = (0,0,1)$ has spherical angles $\theta, \varphi$, which is the direction of the collinear magnetic structure [6]. We assume that $0 \leq \theta \leq \pi, 0 \leq \varphi \leq 2\pi, 0 \leq \psi \leq 2\pi$. in (3). An example of parametrized by the angles $\theta, \varphi, \psi$ local frames, related to transformations $P_i$ (2) are shown in Fig. 5.

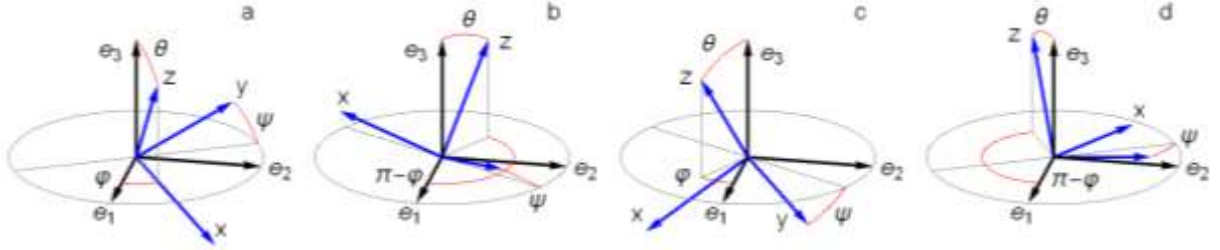

Fig. 5 Example of local coordination frames (black, see eq (2) and (3)) and crystal directions $x, y, z$ (blue) along $a, b$ and $c$ edges of the unit cell, respectively. Numerical values of $\theta, \varphi, \psi$ are 0.601, 0.501, 0.431, respectively. All coordination frames $\boldsymbol{e}_1, \boldsymbol{e}_2, \boldsymbol{e}_3$ have the same handedness. In the space group $Pc2_1n$ the handedness of local environments $Fe_1^{(1)}$ and $Fe_1^{(3)}$ is opposite to that of $Fe_1^{(2)}$ and $Fe_1^{(4)}$, therefore in b) and d) angle $\pi - \varphi$ is shown instead of $\varphi$.

Matrix $R_1 = [\boldsymbol{e}_1, \boldsymbol{e}_2, \boldsymbol{e}_3]$ (1), (3) is equivalent to the Euler matrix of rotation and is a composition of three rotations along axes of Cartesian frame:
$$R_1 = z_1 x_2 z_3 \qquad (4)$$

where
$$z_1 = \begin{bmatrix} \sin\varphi & \cos\varphi & 0 \\ -\cos\varphi & \sin\varphi & 0 \\ 0 & 0 & 1 \end{bmatrix}, \quad x_2 = \begin{bmatrix} 1 & 0 & 0 \\ 0 & \cos\theta & \sin\theta \\ 0 & -\sin\theta & \cos\theta \end{bmatrix}, \quad z_3 = \begin{bmatrix} -\sin\psi & -\cos\psi & 0 \\ \cos\psi & -\sin\psi & 0 \\ 0 & 0 & 1 \end{bmatrix}. \qquad (5)$$

Our considerations are valid for all atomic sites ($Ga_1, Ga_2, Fe_1, Fe_2$) and each site may have its own orientation, e.g, angles $\theta, \varphi, \psi$. An example of considered four local orientations shown in Fig. 5 should not be confused with four possible sites at which iron may be in principle located: $Ga_1, Ga_2, Fe_1, Fe_2$.

    In the first step of analysis we assumed commonly accepted ferrimagnetic structure in which all Fe magnetic moments and thus hyperfine magnetic fields are parallel to the easy magnetization axis. For each component angles $\theta, \varphi, \psi$, (3) defining local orientation of the EFG principal axes, were found to have best description of the all measured spectra, see Fig. 6.

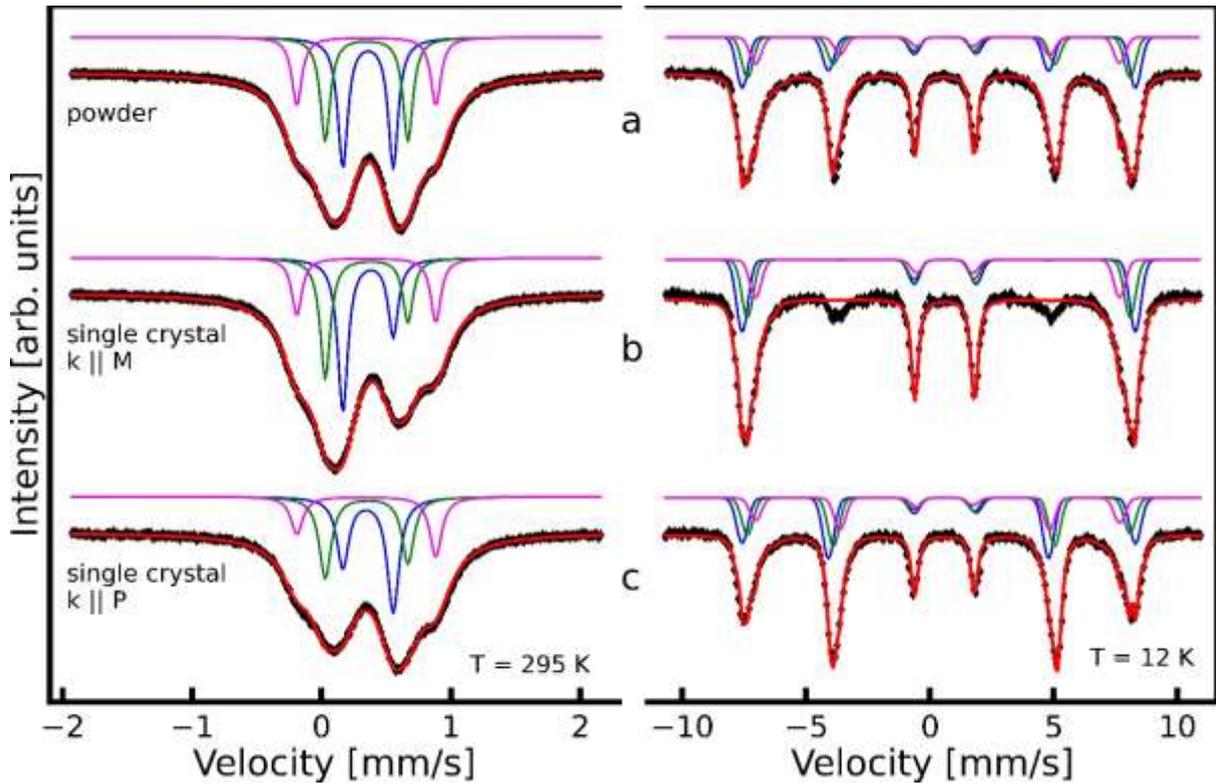

Fig. 6 Measured spectra (black points) and their theoretical description under assumption of ferrimagnetic structure measured on a) powder, b) on sample with the easy magnetic axis and **k** vector o photon parallel and c) on sample with **k** vector of photon parallel to the electrical polarization axis. Lines with blue, green and magenta colours correspond to the iron at sites $Fe_1$, $Fe_2$, $Ga_2$, respectively.

The results of full analysis of the obtained EFG parameters will be published after completion of the measurement with **k** vector of photon perpendicular to the *a* axis in $Pc2_1n$ space group notation. In this report we are considering problem of clear discrepancy between measurement and the theory shown in Fig. 6b, *T*=12 K, where for velocities about -4 and +5 mm/s line intensities do not vanish, showing presence of perpendicular to the direction of gamma rays component.

The spectra shown in Fig. 2 allow to estimate sample orientation. Using approximation of small electrical quadrupole with respect to large magnetic dipole interactions, relative line intensities were estimated for each spectrum as 3:*z*:1:1:*z*:3. For spectra measured with **k** vector perpendicular to the sample plane *a*=0º, *z*= 0.25(4) while for other orientations with *a*=20º shown in Fig. 3: *β*=0º, *z* = 0.50(7); *β*=90º, *z* = 0.46(7); *β*=180º, *z* = 0.61(7). From these data we estimate, that sample is oriented within 1-3º and the best orientation will reduce value of z smaller than its uncertainty. We thus see, that value of *z*= 0.25(4) is not caused by incorrect absorber orientation.

One possibility explaining large *z* value observed in experiments with **k** vector parallel to the easy magnetic axis is formation of domains perpendicular to the easy direction. Quantitative estimation of this issue can be done by comparison of magnetocrystalline and shape anisotropy. The anisotropy constants of $GaFeO_3$ were reported to be $K_a = 1.3 \cdot 10^6$ erg/cc and $K_b = 4.5 \cdot 10^6$ erg/cc [17], and they corresponds to change of energy under rotation of magnetization in *ac* and *bc* plane, respectively. Gain of energy resulting from a shape anisotropy $K_s$ of thin disk, the shape of the absorber, can be estimated as $K_s = \mu_0 M_s^2$, $M_s$ is the spontaneous magnetization and $\mu_0$ vacuum permeability. Since the spontaneous magnetization is 0.67 $\mu_B$/Fe [6] and in our crystal 0.74 $\mu_B$/Fe, it is clear, that magnetocrystalline energies are larger than energy of shape anisotropy for thin disk by about 5 and 20 times, respectively. Therefore, formation of perpendicular domains in *a* o *b* crystal direction is unlikely.

Value $z$ different from zero may indicate for the spin canting or spin disorder and quantitative estimation yields [38]

$$\langle (\boldsymbol{i}_m \cdot \boldsymbol{i}_k)^2 \rangle = \frac{4-z}{4+z}, \qquad (6)$$

where $\boldsymbol{i}_m$, is the unit vector of Fe hyperfine field and $\boldsymbol{i}_k$ is a unit vector of wave vector of photon. From (6) and $z = 0.25(4)$ we get $\langle (\boldsymbol{i}_m \cdot \boldsymbol{i}_k)^2 \rangle = 0.12(1)$, which corresponds to angular distribution of Fe magnetic moment with respect to easy magnetization $c$ axis about 20º.

Finally, we have performed calculations of line intensities and positions under assumption that the magnetic moments are inclined by constant angle from the easy magnetization axis. We have used presented formalism including orientations of the EFG among iron sites. The results are shown in Fig. 1 by lines in colour. The best fit of the lines to the data in Fig. 1 was obtained for inclination of Fe moments from the easy axis by 20º. Full report will be published after completion of measurements in three main crystallographic directions.

## 6. Resume

Floating zone synthesis of stoichiometric $GaFeO_3$ compound and characteristics of single crystals are consistent with results already published [6]. Measurements of single crystal absorbers were presented and the results are also in full agreement with already published data [29], [30]. We show that in case of $\boldsymbol{k}$-vector of photon parallel to the direction of easy magnetization, there is not negligible intensity of $\Delta m = 0$ nuclear transitions and this in turn indicate for magnetic component perpendicular to the magnetization easy axis. We have argued that absorber misalignment as well as the magnetic domains with magnetization perpendicular to the easy axis can-not explain the observed results. The results seem to be in contradiction with commonly accepted ferrimagnetic structure of the ground state. Further works are planned for explanation of the observed properties.

Acknowledgments. This work was partly supported by the National Science Centre, Poland, grant OPUS no 2018/31/B/ST3/00279 and by the Polish Government Plenipotentiary for JINR in Dubna (Project no PWB/168-10/2021).